# Density dependent local structures in InTe phase-change materials

Suyang Sun[1#], Bo Zhang[1#], Xudong Wang[1], Wei Zhang[1,2]*

[1]*Center for Alloy Innovation and Design (CAID), State Key Laboratory for Mechanical Behavior of Materials, Xi'an Jiaotong University, Xi'an 710049, China*
[2]*Pazhou Lab, Pengcheng National Laboratory in Guangzhou, Guangzhou, 510320, China*

[#]These authors contribute equally to this work.
*Corresponding author. Email: wzhang0@mail.xjtu.edu.cn

**ABSTRACT**
Chalcogenide phase-change materials (PCMs) based random access memory (PCRAM) is one of the leading candidates for the development of non-volatile memory and neuro-inspired computing technologies. Recent work shows Indium to be an important alloying element for PCRAM, while a thorough understanding of the parent compound InTe, in particular, its amorphous phase, is still lacking. In this work, we carry out *ab initio* simulations and chemical bonding analyses on amorphous and various crystalline polymorphs of InTe. We reveal that the local geometries are highly density dependent in amorphous structures, forming In-centered tetrahedral motifs under ambient conditions but defective octahedral motifs under pressure, which stems from the bonding characters of its crystalline polymorphs. In addition, our *ab initio* molecular dynamics simulations predict rapid crystallization capability of InTe under pressure. At last, we make a suggestion for better use of Indium and propose an "active" device design to utilize both thermal and mechanical effects for phase-change applications.

# I. INTRODUCTION

The increasing demand for data storage and processing calls for a revolution in current digital memory system.[1] Conventional memory technologies are difficult to balance between programming speed and data retention, resulting in a complex memory hierarchy that consists of fast but volatile static and dynamic random-access memory (SRAM and DRAM) to communicate with the processor and non-volatile but slow memory units, such as solid state drives (SSD) and hard disk drives (HDD), for long-term data storage. Extensive data shuffling between these units in massive scale leads to high power consumption and long delay time. Phase-change materials (PCMs) based random access memory (PCRAM) is one of the emerging non-volatile memories (NVM).[2-8] It has entered the global memory market as storage-class memory (SCM) recently,[9] which fills the performance gap between DRAM and SSD to optimize the overall efficiency of data-centric computing. Moreover, PCRAM is a competitive candidate for neuro-inspired computing[10-18] that unifies computing with storage at the same location.

PCRAM employs large contrast in electrical resistance between the crystalline and amorphous phase of chalcogenides to encode logic states, "1" and "0".[3] The Ge-Sb-Te alloys along the GeTe-$Sb_2Te_3$ pseudo-binary line,[19] in particular, $GeSb_2Te_4$ and $Ge_2Sb_2Te_5$ (GST), are the core materials for PCRAM products. By applying voltage pulses, the local temperature in PCRAM cells can be raised to achieve SET (write) via crystallization (moderate $T$), and RESET (erase) via melt-quenched amorphization (high $T$ and very rapid cooling).[3] High-performance PCRAM calls for robust data storage over 10 years at room $T$, while rapid programming at tens of nanoseconds at elevated $T$. Thanks to the highly fragile nature of PCMs,[20-25] the crystallization time of amorphous PCMs can span over 17 orders of magnitude within a narrow temperature range of 300 °C, making phase-change data storage robust.[25]

In order to extend the capacity of PCRAM towards cache-type[26-28] or embedded memory[29-31] applications, doping and alloying are frequently employed to accelerate the crystallization speed at elevated $T$, while to increase the crystallization $T$ of the amorphous phase at the same time. In our recent work, we focused on doping $Sb_2Te_3$ with Sc, forming $Sc_{0.2}Sb_{1.8}Te_3$ (SST), which leads to an ultrafast SET speed of ~0.7 ns without any pre-programming operation.[32] Under the same condition, the minimum SET time for GST devices is ~10 ns. Yet, the crystallization temperature $T_x$ and the 10-year data retention temperature $T_d$ of SST (170 °C, 87 °C) are still comparable to those of GST (150 °C, 82 °C).[32] The underlying mechanisms were thoroughly discussed in Refs.[32-38]. Nevertheless, the thermal stability of SST is not yet suitable for embedded memory, which calls for higher $T_x$ and $T_d$ to cope with the soldering process.[29]

Recently, Indium doped GST alloys[39,40] (in particular, $In_{0.9}Ge_2Sb_2Te_5$, short as IGST) were shown to have excellent thermal stability ($T_x$=270 °C, $T_d$=180 °C) and good operation speed (6 ns). In addition, Indium has long been used as a key ingredient in important PCMs, including AgInSbTe (AIST),[41-43] $In_3SbTe_2$ (IST),[44-49] and $InGeTe_2$ (IGT) alloys.[50,51]

However, the parent phase, InTe, in particular, its amorphous phase, remains elusive. In this work, we present a comprehensive theoretical work of InTe based on *ab initio* simulations. We thoroughly discuss the structural details and bonding characteristics of the crystalline polymorphs and the amorphous phase of InTe, which clearly explain the strong dependence of In-bonded motifs on mass density. These results help guide the design of structure-sensitive PCMs for practical applications.

## II. METHODS

We performed density functional theory (DFT) calculations and DFT-based *ab initio* molecular dynamics (AIMD) simulations using the second-generation Car-Parrinello molecular dynamics scheme[52] implemented in the CP2K package.[53] We used the Perdew-Burke-Ernzerhof (PBE) functional[54] in combination with the Goedecker pseudopotential.[55] The time step was set to 2 fs. The electronic structure calculations and chemical bonding analyses were done by using the Vienna Ab-initio Simulation Package (VASP) code[56] in combination with the Local Orbital Basis Suite Toward Electronic-Structure Reconstruction (LOBSTER) code,[57-59] where the projector augmented wave (PAW) method[60] and PBE functionals were used with an energy cutoff of 450 eV. The Brillouin zone of TlSe-, NaCl-, CsCl-type InTe crystals was sampled by k-point mesh ≥4×4×4, while the amorphous supercell models were sampled with gamma point only. Van der Waals corrections[61] were included in all DFT and AIMD calculations. The atomic structures were visualized by the VESTA software.[62]

## III. RESULTS AND DISCUSSION

Three crystalline polymorphs are found for InTe, namely, the tetragonal (*t*-) phase (TlSe-type, space group *I4/mcm*) under ambient conditions, and the rock-salt (*rs*-) phase (NaCl-type, space group $Fm\bar{3}m$) and the cubic (*c*-) phase (CsCl-type, space group $Pm\bar{3}m$) under high pressure (~5 GPa and ~15 GPa, respectively).[63] The DFT-relaxed structures of the three phases (zero pressure) are presented in Figure 1. The calculated lattice parameters of *t*-InTe unit cell are 8.66, 8.66 and 7.30 Å, which are in good agreement with experimental values.[64] The *t*-phase unit cell (Figure 1a) contains edge-shared In-centered tetrahedral motifs made of short In–Te bonds (2.86 Å) and interstitial-like In atoms that are weakly coupled to the surrounding atoms with long In–In (3.65 Å) and In–Te (3.67 Å) contacts. The *t*-phase unit cell can be viewed as a distorted 2×2×2 CsCl-type supercell (Figure 1c) with Te atoms shifting from ideal position to form edge-shared InTe$_4$ tetrahedron chains, leaving other In atoms being weakly coupled. The calculated equilibrium and high-pressure lattice parameters of *rs*-InTe are 6.24 Å and 6.05 Å (5 GPa), while those of *c*-InTe are 3.84 Å and 3.57 Å (15 GPa). It is interesting to note that *rs*-InTe can be also obtained in nano-crystalline form by mechanical milling of equiatomic mixtures of elemental In and Te powders over 2-40 hours, which shows an enlarged lattice parameter[64] of ~6.13 Å with respect to the high-pressure value ~6.0 Å at 5 GPa.[63] The nano-crystalline *rs*-InTe was shown to be stable over years.[64] In fact, the cohesive energy values of *t*-InTe (–0.313 eV/atom) and *rs*-InTe (–0.310 eV/atom) are very close in the bulk form at zero pressure condition. The lattice expansion upon removal of external pressure also agrees well with our DFT calculations. This mixture of tetrahedral and octahedral bonding configurations at ambient conditions and low pressure is a unique feature of InTe, as tetrahedral bonds are found dominant in other Indium Telluride binary compounds, such as $In_2Te_3$[65,66] and $In_2Te_5$,[67] which are also claimed to be useful PCMs. The theoretical mass density values of crystalline polymorphs of InTe are 5.88 g/cm$^3$ for the *t*-phase, 6.64 and 7.26 g/cm$^3$ for the *rs*-phase without and with 5 GPa pressure, and 14.22 and 17.76 g/cm$^3$ for the *c*-phase without and with 15 GPa pressure. In the following, we refer to these theoretical values, if not specified otherwise.

Our electronic structure calculations show a semiconducting behavior for *t*-InTe, while metallic characters for the other two phases with and without pressure (this observation remains valid for hybrid functional[68] DFT calculations), as shown by the corresponding density of states (DOS) and band structures in Figure 1 and Figure S1. The crystal orbital overlap population (COOP) can reveal the covalent interaction by distinguishing bonding (stabilizing, positive) contributions from antibonding (destabilizing, negative) contributions to the electronic structure, which can directly be extracted using DFT plane-wave functions.[57-59] As shown in Figure 1a, there is no antibonding interaction at the Fermi level $E_F$, indicating good chemical stability of the model, in line with our previous work.[47] Regarding *rs*-InTe, the valence shells of both In ($5s^25p^1$) and Te ($5s^25p^4$) atoms are not fully occupied, and the average number of *p* electrons is 2.5, giving rise to a metallic character. Note that in (distorted) rock-salt GeTe and GST, where the average number of *p* electrons per site equals 3, a special covalent bonding mechanism—metavalent bonding (MVB)—prevails,[69-72] making these solids semiconductors instead of metals. Although strong antibonding interactions are found at the Fermi level for both *rs*- and *c*-InTe, these two phases can exist under pressure.

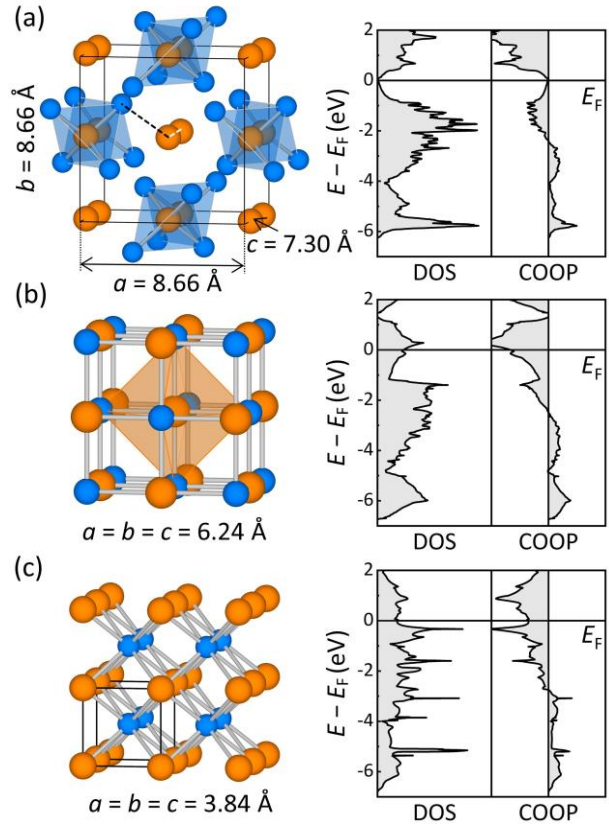

**Figure 1.** (a)-(c) The DFT-relaxed atomic structures, the calculated DOS and COOP curves of *t*-, *rs*- and *c*-InTe crystals under zero pressure. In and Te atoms are rendered with orange and blue spheres, respectively. The shortest interatomic contacts for interstitial-like In atoms in the *t*-InTe are highlighted with black and white dash lines.

The amorphous (*a*-) models of InTe compound were obtained by AIMD calculations in the NVT scheme. Starting from a 3×3×3 *rs*-InTe supercell with 216 atoms, the model was fully randomized above 2500K for 15 ps, and was quenched down to 1200K and equilibrated there for 30 ps. An amorphous model was then generated by quenching the liquid down to 300K in 72 ps. During this quenching process, the simulation was stopped after every 100 K, and the box size was adjusted to reduce the internal stress. After 30 ps annealing at 300 K, the model was quenched down to zero K for electronic structure and chemical bonding calculations. The same procedure was repeated twice for statistics. The average lattice edge is determined to be 19.56 Å, corresponding to a mass density of $D_a$ = 5.81 g/cm$^3$ for *a*-InTe. This value is very close to that of *t*-InTe, 5.88 g/cm$^3$. In addition, we simulated three additional *a*-InTe models with higher density, $D_{rs}$ = 6.64 g/cm$^3$ (lattice edge 18.71 Å), corresponding to that of zero-pressure *rs*-InTe. The internal pressure is calculated to be ~2.1 GPa on average. This set of models could be relevant to practical use, as the programming volumes are highly confined in PCRAM devices.

The atomic structures, DOS and COOP plots of the two sets of *a*-InTe are shown in Figure 2a and Figure S2. Narrow band gaps are consistently found in *a*-InTe ($D_a$), while only pseudo gaps are found in *a*-InTe ($D_{rs}$). This trend holds for hybrid functional calculations as well (Figure S3). Similar to the crystalline phases, the finite electronic states lead to antibonding interactions at the Fermi level in a-InTe ($D_{rs}$), see Figure 2b. The projected COOP (pCOOP) analysis reveals In–Te bonds to be the major contribution of antibonding interaction at and/or around the $E_F$ in *a*-InTe of both densities. Regarding the electrostatic interaction, we made Mülliken charge analysis using the LOBSTER code. Similar to other PCMs,[73-76] In atoms transfer small portion of electrons to Te atoms, making all atoms slightly charged. The degree of charge transfer in *a*-InTe gets reduced when subjected to pressure (Figure 2c). This trend is in line with its crystalline counterparts, i.e. the Mülliken charges are (+0.24, –0.24) and (+0.1, –0.1) for the *t*- and *rs*-phase of InTe, respectively.

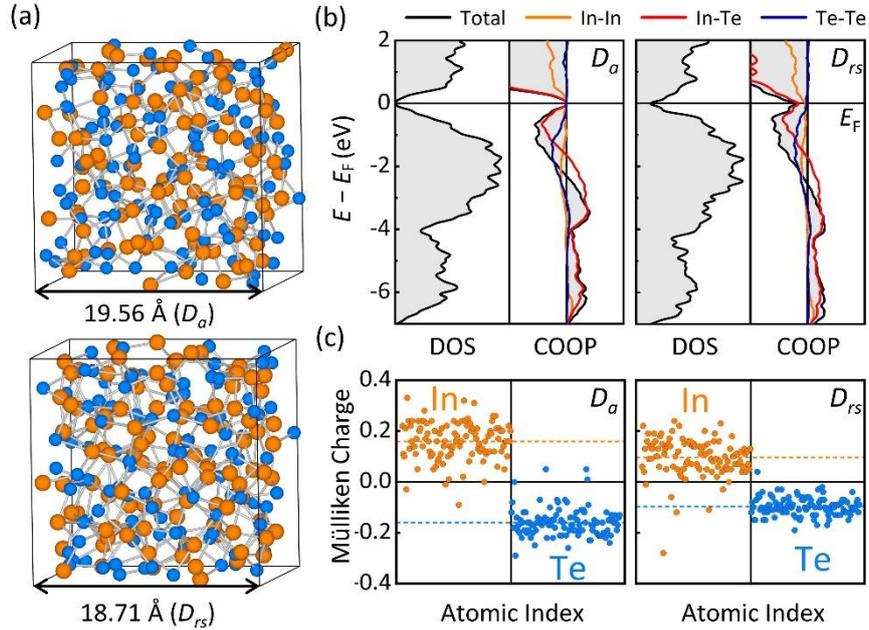

**Figure 2.** (a) The atomic structures of *a*-InTe at both $D_a$=5.81 g/cm$^3$ and $D_{rs}$=6.64 g/cm$^3$ obtained by AIMD simulations. (b-c) The calculated DOS/COOP curves and Mülliken charges of the two amorphous models shown in (a). The projected COOP curves for In–In, In–Te and Te–Te contacts are plotted in orange, red and blue, respectively. The orange and blue dashed lines mark the average Mülliken charge values of In and Te atoms, respectively.

Figure 3 shows the radial distribution functions (RDFs) of *a*-InTe, which were calculated using the AIMD trajectories equilibrated at 300 K. Three models were considered for each density. The first peak position of In–In and In–Te contacts of the low-density *a*-InTe ($D_a$) is at 2.87 Å and 2.93 Å, respectively, which gets reduced to 2.79 Å and 2.86 Å as density increases to $D_{rs}$. The bond population $B_{AB}$ can be calculated by integrating the projected COOP(*E*) on a specific pair of atoms A and B up to $E_F$, viz. $B_{AB} = \int_{-\infty}^{E_F} \text{COOP}_{AB}(E)dE$, which characterizes the strength of chemical bonds. The overall behavior of $B_{AB}$ distribution is similar to the case in amorphous GeTe and GST,[31,77] as bonding interactions are mostly found for In–In and In–Te contacts but antibonding interactions for Te–Te contacts. The difference is that the In–In bonds show slightly larger bond population values than the In–Te bonds below 3.5 Å, while those of Ge–Ge and Ge–Te bonds mostly overlap in *a*-GeTe and *a*-GST,[31,77] indicating that the homopolar In–In bonds could potentially be more difficult to break down in *a*-InTe. By multiplying $B_{AB}$ and RDF at a given interatomic distance, the bond-weighted distribution function, $\mathrm{BWDF} = \sum_{B>A}[\delta(r - |\boldsymbol{r}_{AB}|)] \times B_{AB}$ shows a clear crossover from bonding to antibonding interaction as interatomic distance $r$ increases. Despite the difference in density, similar crossover values are found in the two sets of *a*-InTe models, which are regarded as cutoff values, i.e. 3.37 Å (In–In), 3.43 Å (In–Te) and 3.03 Å (Te–Te), for the following structural analyses.

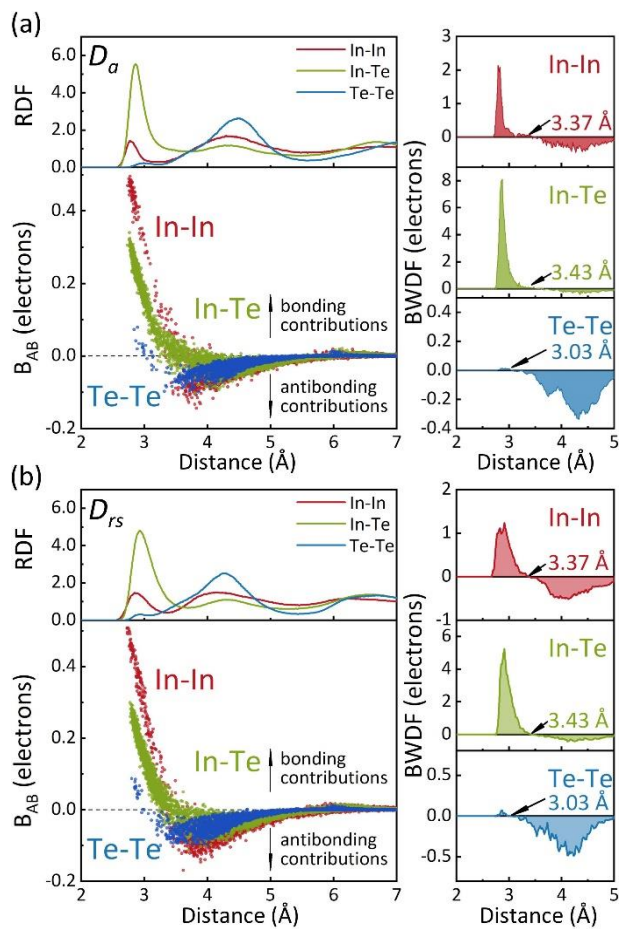

**Figure 3.** The partial RDF, bond population $B_{AB}$, and BWDF of *a*-InTe of two densities, (a) $D_a$ and (b) $D_{rs}$. The In–In, In–Te and Te–Te RDF curves are plotted in red, green and blue respectively. For each density, bond population data of contacts up to 7 Å were collected. The crossover position from bonding to antibonding interactions in BWDF curves can be regarded as cutoffs (In–In: 3.37 Å; In–Te: 3.43 Å; Te–Te: 3.03 Å) of chemical bonds.

As displayed in Figure 4, the angular distribution function (ADF) shows that the peak position of In-centered motifs in *a*-InTe is at ~103° at equilibrium but ~90° under pressure. As density increases, the average coordination number (CN) increases from 4 to 5, suggesting a change from tetrahedral (tetra-) to octahedral (octa-) motifs for In atoms. Regarding Te atoms, both peaks in ADF are found around 90°, and the average CN changes from 3 to 4, indicating a change from 3-fold to 4-fold defective octahedral motifs. Clearly, there is a strong dependence of local geometries on mass density. To quantify this structural difference, we computed the bond order parameter,[78] $q = 1 - \frac{3}{8}\sum_{X>Y}\left(\frac{1}{3} + cos\,\theta_{XAY}\right)^2$ where $\theta_{XAY}$ represents the bond angle of the center atom A with its two neighboring atoms X and Y. The bond order analysis confirms a high concentration of tetra-In atoms over ~58% at $D_a$ but a much lower value ~20% at $D_{rs}$. Other In atoms and Te atoms are mostly found in defective octahedral coordination with varying neighbors from 3 to 6 at both densities. The raised number of octa-bonds at $D_{rs}$ also largely increases ratio of 4-fold primitive rings (Figure 4c), which could potentially enhance the probability of nucleation at elevated temperatures.[14,79-84]

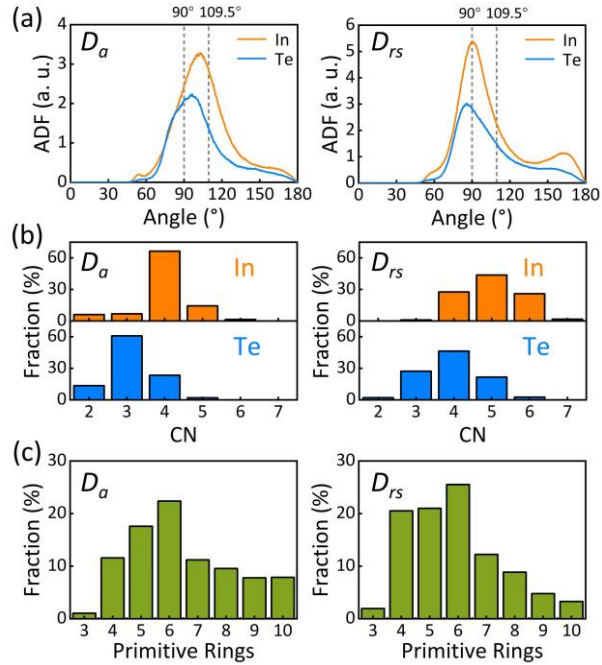

**Figure 4.** The (a) ADF, (b) CN distribution and (c) primitive rings statistics of *a*-InTe of two densities. The bond angle of ideal octahedral (90°) and tetrahedral (109.5°) motifs are marked by dash lines in (a).

In amorphous GeTe and GST, the concentration of tetra-Ge atoms is ~25–30%, and nearly all the tetra-Ge motifs contain at least one homopolar Ge–Ge bond.[81-83] The crucial role of homopolar Ge–Ge bonds in stabilizing tetra-Ge units locally was quantified by thorough pCOOP analysis in Ref. [85]. Here, we carry out similar analysis on *a*-InTe. In stark contrast with *a*-GeTe, there is no clearly dependence of homopolar bonds and tetrahedral motifs. As shown in Figure 5, there are significant amounts of tetra-In motifs bonded with pure heteropolar In–Te bonds at both densities, i.e., 46.4% ($D_a$) and 37.3% ($D_{rs}$) tetra-In[In$_0$Te$_4$] with respect to the total number of tetra-In motifs at each density. Moreover, there is not much antibonding interaction for these heteropolar-bonded motifs at the Fermi level and below, and the same holds for the tetra-In motifs with homopolar In–In bonds. Mass density only changes the ratio of tetrahedral motifs but not its dependence on homopolar bonds. This behavior stems from the respective crystalline phase. For GeTe and GST, there is no crystalline phase that can sustain tetrahedral coordination experimentally. If GeTe takes the hypothetical TlSe structure, DFT calculations show a highly unfavorable configuration, which is ~265 meV/atom higher than its ground state rhombohedral phase, and even higher than its amorphous phase by ~126 meV/atom. This hypothetical GeTe quickly collapses upon DFT relaxation in supercells. But for crystalline InTe, both tetrahedral and octahedral motifs can exist at ambient conditions, and all tetra-In atoms are heteropolar-bonded in $t$-InTe. This change in tetrahedral motif ratio upon density variation is highly consistent with that observed in amorphous IST.[46] We note that this density-structure sensitivity in IST originates mainly from its parent alloy InTe, but much less from the other parent alloy InSb, as the latter shows no strong density-structure dependence in its amorphous phase.[86]

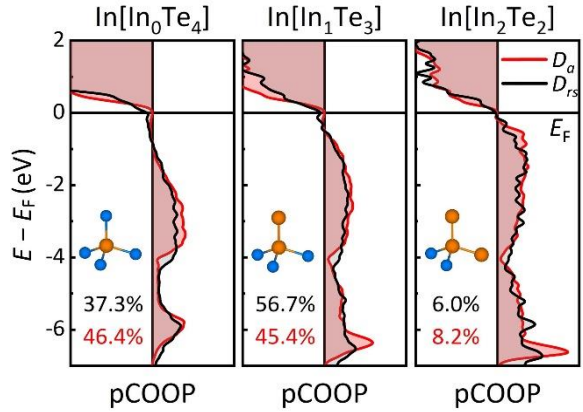

**Figure 5.** The pCOOP curves for In-centered tetrahedral motifs. The pCOOP curves were grouped into tetrahedral motifs with all heteropolar In–Te bonds, and with one, and two homopolar In–In bonds. The respective motifs are sketched as insets.

Finally, we investigate the crystallization kinetics of $a$-InTe. In practice, PCRAM devices undergo long-term thermal annealing during the fabrication process, and the initial state is usually in the SET state. To RESET the memory cell, an amorphous area is created on top of the bottom electrode contact (BEC), and is surrounded by the crystalline matrix. For the subsequent SET operation, the amorphous region can be crystallized through interfacial growth via the boundary and/or nucleation and grain growth, see Figure 6a. We carried out AIMD calculations at 600 K by heating up two $a$-InTe models with different densities, and after 500 ps, no clear sign of crystallization was observed for both calculations. Although the ratio of 4-fold rings is increased at higher density $D_{rs}$, the abundance of five- and six-fold rings adds complexity to the incubation process in $a$-InTe, in contrast to the case of $a$-GST[79-82] and $a$-SST[28,33] where 4-fold rings dominate. The initial configuration and the one after 500 ps of $a$-InTe ($D_{rs}$) are shown in Figure 6b. Then we constructed two additional models with crystalline boundary to assess the crystal growth tendency. Two (100) layers of $rs$-InTe were added to $a$-InTe models along one axis where the periodic boundary condition was removed. The lattice edge of the $a$-InTe ($D_{rs}$) model fits perfectly to the seed layers, while for the $a$-InTe ($D_a$) model, the simulation box was slightly adjusted to remove the lattice mismatch. We chose an optimal distance between the seed layers and the amorphous slab, guaranteeing the shortest interatomic distance of any two atoms from the two regions higher than the typical bond length in *a*-InTe ~3.0 Å. After annealing at 300 K over 3 ps to stabilize the total energy, we moved the seed layers towards the amorphous slab six times to reduce the lattice edge, reaching the density $D_a$ and $D_{rs}$ for the amorphous slab, respectively. At each interval, the models were equilibrated over 3 ps. The two *a*-InTe models with crystalline boundary were then heated to 600 K. At higher density $D_{rs}$, crystal growth proceeds quickly from the boundary, and gets completed after 140 ps, as shown in Figure 6c. While the growth is still very limited for the other model at lower density $D_a$ after 500 ps.

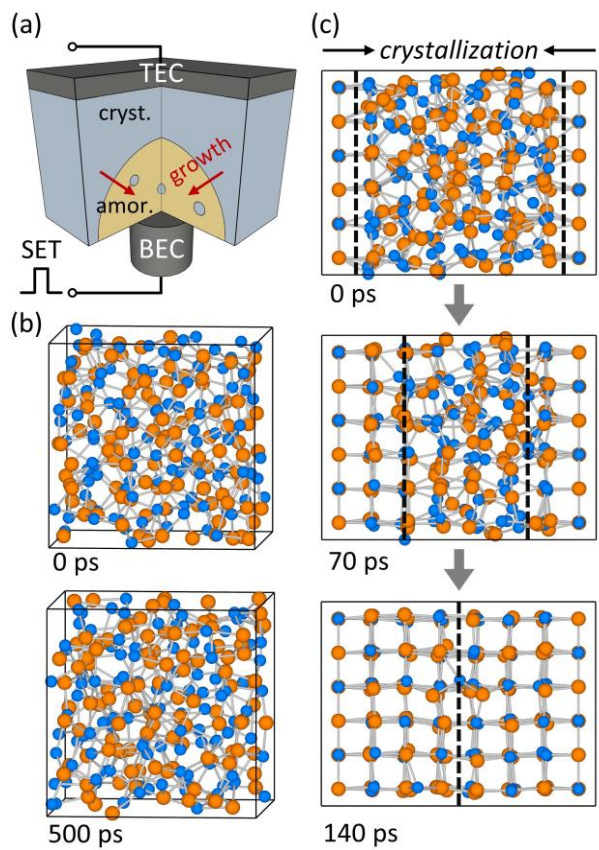

**Figure 6.** (a) The sketch of a typical mushroom-type PCRAM cell. TEC and BEC are the abbreviations of top and bottom electrode contact. When a SET pulse is sent to the device, the amorphous region can be crystallization via crystal growth from the boundary or nucleation and grain growth from inside. (b) Snapshots of an *a*-InTe ($D_{rs}$) model annealed at 600K over 500ps. (c) Snapshots of *a*-InTe ($D_{rs}$) model with crystalline boundaries annealed at 600K at 0, 70 and 140 ps.

Our AIMD crystallization simulations suggest that crystallization speed of *a*-InTe is highly dependent on the density of amorphous structures. Higher density, and thus fewer tetrahedral motifs and more four-fold rings in *a*-InTe, would facilitate crystallization in SET process of devices, which, however, would weaken the amorphous stability. As reported in Ref. [40], the InTe based devices could reach a quite high SET speed ~10 ns, but show a much lower $T_x$ ~187 °C, as compared to that of free-standing InTe thin films, $T_x$ ~300 °C.[87] A plausible explanation to account for this large difference in $T_x$ is that the *a*-InTe region in devices is subjected to pressure imposed by the surrounding crystalline matrix and isolation layers, resulting in a rapid crystallization process into the metastable *rs*-phase. While in free-standing thin films, *a*-InTe shows a much higher barrier for crystallization, and the obtained crystallized state is in the ground-state *t*-phase. Therefore, it is important to monitor the thermal and mechanical conditions during device fabrications, when InTe, IST and other highly sensitive materials are exploited as the core materials. The difference in programming volume could possibly lead to large variances in device performance, in particular, thermal stability and SET speed.

Our findings also provide better understanding of the performance of IGST devices.[39] As stated before, small amount of In content incorporated in GST could shorten the SET time from ~30 ns to ~6 ns, and the In atoms were demonstrated to form octahedral bonds with Te atoms in crystallized *rs*-IGST.[39] This observation can be attributed to the fact that In–Te bonds are highly compressed inside the GST crystalline matrix, as the lattice parameter of *rs*-GST (6.03 Å) coincides with that of *rs*-InTe (6.0 Å) under 5 GPa in experiments.[63] The preferentially formed octahedral In–Te bonds in IGST could assist the incubation process for faster SET speed.[39] However, we also note that too much Indium content should be avoided, because if InTe takes the major fraction in IGST, the extra higher-fold primitive rings may appear, adding complexities to the formation of *rs*-phase seeds. Regarding the use of Indium in other related PCMs with larger lattice parameters, such as SnTe, PbTe, Sn-Sb-Te, Pb-Sb-Te and others,[88-90] the In–Te bonds are no longer subjected to compressive strain, therefore, In atoms should be mostly tetrahedrally coordinated in the amorphous phase, serving mainly as structural barriers for crystallization. This alloying scheme could be useful for the development of embedded memories[29-31] and optical modulators,[91] which emphasize more on superior amorphous stability rather than rapid crystallization speed.

**IV. CONCLUSION**
In summary, we have carried out comprehensive *ab initio* investigations to gain in-depth understanding of the atomistic details and bonding characteristics of InTe. We found that the In-bonded motifs in the amorphous structures are highly density dependent, namely, octahedral motifs dominate in compressed amorphous structures, while tetrahedral ones prevail in amorphous structures under ambient conditions or under low pressure. This trend can be understood from its crystalline polymorphs of rock-salt and tetragonal phase, and the very small energy difference between the two phases of few meV per atom makes Indium a multi-functional element for PCM applications. Moreover, our findings on the density-structure sensitivity point towards the design of an "active" device setup that combines structure-sensitive PCMs with stress actuators or modulators, which could tailor the crystallization capability via external mechanical forces to achieve both better data retention and faster operation speed. In fact, stress-induced phenomena in PCMs have been extensively investigated in literature.[92-97] Our work shall serve as a stimulus for further development of phase-change devices that utilize both thermal- and mechanical-induced transitions.

**ACKNOWLEDGEMENTS**

W. Z. acknowledges the National Natural Science Foundation of China (Grant No. 61774123), 111 project 2.0 (BP2018008) and the International Joint Laboratory for Micro/Nano Manufacturing and Measurement Technologies of XJTU. The authors acknowledge the computational resource provided by the HPC platform at XJTU.